\documentclass[%
 reprint,
 amsmath,amssymb,
 aps,prl
]{revtex4-1}

\usepackage{graphicx}
\usepackage{dcolumn}
\usepackage{bm}
\usepackage[colorlinks]{hyperref}
\usepackage[utf8x]{inputenc}
\usepackage{braket}
\usepackage{amssymb}
\usepackage{amsmath}
\usepackage{graphicx}
\usepackage[colorinlistoftodos]{todonotes}
\usepackage{lipsum}

\begin{document}

\preprint{APS/123-QED}

\title{Conservation of Quantum Correlations in Multimode Systems with 
${U(1)}$ Symmetry}

\author{G.Buonaiuto}
\email{gbuonaiuto1@sheffield.ac.uk}
\author{D.M. Whittaker}%
\author{E.Cancellieri}
\email{New Address:Department of Physics, Lancaster University, Lancaster, LA1 4YB, United Kingdom}

\affiliation{Department of Physics and Astronomy, University of Sheffield, Sheffield S3 7RH, United Kingdom}%


\begin{abstract}
We present a theoretical investigation of the properties of
quantum correlation functions in multimode system. We define a total $m^{th}$ order
equal-time correlation function, summed over all modes, which is shown to be conserved if the Hamiltonian possesses U(1) symmetry. It is also conserved in the presence of dissipation, provided the loss rate is the 
same for all modes of the system. As examples, we demonstrate this conservation using numerical simulations of a coupled cavity system and the Jaynes-Cummings 
model.
\end{abstract}

\pacs{Valid PACS appear here}
\maketitle


In this letter, we investigate theoretically the properties of
number operator correlation functions in a multimode system, where the
Hamiltonian has $U(1)$ symmetry \cite{Isham}, or equivalently commutes with the
total number operator for all the modes in the system. We define a total
$m^{th}$ order equal-time correlation function for the system, summed over all modes. We show that this quantity is conserved in a closed system, and also in a dissipative system if all the modes have linear gain or losses of the same magnitude. As examples, we demonstrate this conservation
using numerical simulations of a coupled cavity system and the Jaynes-Cummings
model. We show that the second order correlation functions for individual
modes are not conserved, but the total correlation function is constant.
The link between the conservation of intensity correlation functions and the $U(1)$ invariance, opens the way to use photon counting measurements to determine when the system is undergoing a spontaneous global $U(1)$ symmetry breaking, with broad applications in condensed matter systems, like quantum Hall bilayers \cite{Pik} and spin liquid in Kagome lattice \cite{Carr}.

Equal time correlation functions, particularly the second order, equal time,
correlation function (SOC), $g^{(2)}(t,t)$ (often referred to as $g^{(2)}(0)$
in a steady state), are important tools, both theoretically and experimentally
for investigating the statistical properties of fluctuations in a quantum
system \cite{David}. The SOC characterizes the variance in the occupation of a mode. It is
defined in such a way that for a classical field $g^{(2)}(0) \ge 1$, so that a
value less than one is regarded as a clear signature of `quantum' behavior,
while the smallness of $g^{(2)}(0)$ is an important figure of merit for a
single photon source. Second, and higher order correlation functions are also
relevant in a range of photon counting experiments, such as the Hanbury Brown
- Twiss interferometer  \cite{hau2}. Understanding the dynamics of correlation functions is
important in the investigation of phase transitions, entanglement and other
quantum properties of optical and condensed matter systems.

In this letter, we investigate the properties of quantum correlation functions
in a multimode system. In this case there are numerous
correlation functions, describing the fluctuations in each mode individually,
but also the cross-correlations between different modes. We define a total,
$m^{th}$-order, correlation function for such a system, summing over all the
different modes. We show that this total correlation function is conserved if
the Hamiltonian governing the dynamics exhibits $U(1)$ symmetry and,in a dissipative scenario, if all the modes experience the same dissipation
with Lindbald terms\citep{Petruccione}, which are linear in
the system operators. A Hamiltonian which possesses $U(1)$ symmetry will
preserve the total number of excitations in a non-dissipative system;
dissipation will, of course, mean that the number is not conserved.

\textit{The Total Correlation Function.} For a multimode system, there are
multiple second order quantum correlation functions corresponding to
the different modes which are measured. At  equal 
times, these have the form
\begin{equation}
\label{socqc}
g^{(2)}_{ij}(t,t)=\frac{\braket{c_i^{\dagger}c_j^{\dagger}c_i c_j}}
{{\braket{c_i^{\dagger}c_i}\braket{c_j^{\dagger}c_j}}},
\end{equation}
where $c^{\dagger}_i$, $c_i$ and $c^{\dagger}_j$, $c_j$ are the creation and
annihilation operators for modes $i$ and $j$, which may be bosonic or
fermionic, and $\braket{...}$ is the time dependent mean value of the operators. If $i=j$, Eq. \eqref{socqc} quantifies the fluctuations within a mode, otherwise
it describes correlations between the fluctuations in different modes. This
definition can be extended to higher order correlations: the $m^{ th}$ order 
function contains $m$\/ creation and annihilation operators, potentially for
$m$\/ different modes.

We define the {\em total $m^{th}$ order correlation function} for the output field of a multimode
system as
\begin{equation}
g_{tot}^{(m)}(t,t)=\frac{\braket{J}}{\braket{N}^{m}},
\end{equation}
where
\begin{equation}
\braket{J}=\braket{:(\sum_{i}c_{i}^{\dagger}c_{i})^{m}:}
\end{equation}
and
\begin{equation}
\braket{N}=\braket{\sum_{i} c_{i}^{\dagger}c_{i}}
\end{equation}
is the total number operator for the system. $\braket{:..:}$ indicates the
normal ordering of the enclosed operators, and the sum extended over every mode in the system. Note that this is not the same as
summing the $g_{i,j}^{(m)}$ over all the modes because every term is normalized by
the $m^{th}$ power of the total occupation of the system. However, we regard
our definition as a more natural sum, because it weights the contribution of
each mode according to its occupation. Furthermore, in a system where all the
modes are photonic with equal external coupling, it represents the $m^{th}$
order correlation function which would be obtained if the total emission were
to be measured, without resolving the individual modes. In the case of a
single mode and $m=2$\/ it is identical to the standard SOC, given by 
Eq.(\ref{socqc}) with $i=j$.

\textit{Conservation of $g_{tot}^{(m)}(t,t)$ in closed system}.
In order to determine the condition that the total correlation function is 
conserved, we 
equate its time derivative to zero,  giving
\begin{equation}
\label{conditcl}
\left(\frac{d}{dt}\braket{J}\right)\braket{N}^{m}-\left(m\braket{N}^{m-1}\frac{d}{dt}\braket{N}\right)\braket{J}=0.
\end{equation}
For closed system that is guaranteed by: 

\begin{align}
\label{conditcl1}
   \frac{d}{dt}\braket{N}=0
   ,\hspace{0.3cm}\frac{d}{dt}\braket{J}=0,
\end{align}
which means that the Hamiltonian must commute with both $N$ and $J$. Now, the normally ordered operator $J$ can be written as a power
expansion in the total number operator, $J=\sum_{k}^{m}d_{k}N^{k}$, where
$d_{k}$ are numerical coefficient related to the order of the expansion
\cite{VARGAS}. The condition $[H,J]=0$ is thus satisfied provided $[H,N]=0$,
so this is our only requirement. The condition $[H,N]=0$ is equivalent to
invariance under the gauge transformation 
\begin{equation}
\label{condit2}
   \begin{cases}
   \tilde{c_{i}}^{\dagger}=c_{i}^{\dagger}e^{i\phi}\\\tilde{c_{i}}=c_{i}e^{-i\phi} &\forall i 
   \end{cases}\nonumber
\end{equation}
which corresponds to a global $U(1)$ symmetry of the Hamiltonian $H$. 

This result can be summarized in a general theorem:
\textbf{Theorem:}\textit{ For any closed system with an arbitrary number
of modes, described by Hamiltonian $H$, iff $H$ globally possess a $U(1)$ symmetry, i.e. 
$[H,N]=0$, where $N$ is the
total number operator, then the total $m^{th}$ order equal time 
correlation function
$g^{(m)}_{tot}(t,t)$ is a conserved quantity.}

\textit{Two modes quantum fields}. To underline the link between the global
$U(1)$ symmetry of the Hamiltonian and the conservation of the SOC function,
we first consider the case of two closed, undriven linear bosonic modes, coherently
coupled. 
This system can be described by a Hamiltonian of the form
\begin{equation}
\label{Htm}
H_{tm}=\omega_{1}a_{1}^{\dagger}a_{1}+\omega_{2}a_{2}^{\dagger}a_{2}+
\tau(a_{1}^{\dagger}a_{2}+a_{2}^{\dagger}a_{1}),
\end{equation}
where $\omega_{i}$ are the energies of each mode and $\tau$ is the coupling
strength between modes. Note that we can add non-linear terms with no effect provided they commute with the number operators for each mode, so this discussion could equally apply to a two-mode Bose-Hubbard model. 

We need to evaluate the time derivative for three correlators:
the two auto-correlation functions $(g_{1}^{(2)},g_{2}^{(2)})$ and the cross
correlation function $(g_{1,2}^{(2)}=g_{2,1}^{(2)})$. It is easy to show the dependence of the auto-correlation,
$\braket{a_{1}^{\dagger}a_{1}^{\dagger}a_{1}a_{1}}$, from the hopping terms:
\begin{eqnarray}
\label{twomodeG}
\frac{d}{dt}\braket{a_{1}^{\dagger}a_{1}^{\dagger}a_{1}a_{1}}
&=&  
-i\tau (2\braket{a_{1}^{\dagger}a_{1}a_{1}a_{2}^{\dagger}}
-2\braket{a_{1}^{\dagger}a_{1}^{\dagger}a_{1}a_{2}} 
+\braket{a_{1}^{\dagger}a_{2}}
\nonumber\\  
& &
-\braket{a_{2}^{\dagger}a_{1}}), 
\end{eqnarray}
while the derivative of the number of particles for mode 1, $\braket{n_{1}}$, is
\begin{equation}
\label{twomodeN}
\frac{d}{dt}\braket{n_{1}}=-i\tau
\braket{a^{\dagger}_{1}a_{2}-a^{\dagger}_{2}a_{1}},
\end{equation}
with an equivalent form for mode 2.
The autocorrelation function for mode 1 (and
analogously for mode 2) is not conserved. 
However,
considering now our \textbf{Theorem}, we note that though the Hamiltonian does
not commute with the number operator for each mode ($[H_{tm},n_{1,2}]\neq 0$),
i.e. the $U(1)$ symmetry is locally broken, it does commute with the total
number operator $N=n_{1}+n_{2}$. This is a consequence of the global
$U(1)$ symmetry of the Hamiltonian. Therefore, it can be expected that total
SOC function, $g_{tot}^{(m)}$ is conserved. This can be confirmed by evaluating the time derivative of the cross correlation between the two modes. the expression contains terms which precisely cancel the derivative of the auto-correlation function:

\begin{eqnarray}
\label{twomodecross}
\frac{d}{dt}\braket{a_{2}^{\dagger}a_{2}a_{1}^{\dagger}a_{1}}&=& i\tau (
\braket{a_{2}^{\dagger}a_{2}a_{1}^{\dagger}a_{2}}+
\braket{a_{2}^{\dagger}a_{1}a_{1}^{\dagger}a_{1}}-
\nonumber \\
& &
\braket{a_{1}^{\dagger}a_{2}a_{1}^{\dagger}a_{1}}-
\braket{a_{2}^{\dagger}a_{2}a_{2}^{\dagger}a_{1}}). 
\end{eqnarray}
Increasing the number of modes, the above argument 
encompasses the 1D Bose-Hubbard Hamiltonian. Moreover, it can be also extended
to bosonic quantum networks \cite{Baier}, which can describe a huge variety of
physical systems, like photonic or polaritonic lattices (Lieb \cite{Charles},
Kagome \cite{Endo} and Graphene \cite{Jacqmin}), quantum networks and
collective phenomena involving parametric processes (like parametric up/down
conversion).

   

\textit{Jaynes-Cummings Hamiltonian}. It is also interesting to
consider the case of mixed bosonic and fermionic systems, for example, the
Jaynes-Cummings-Hubbard model \cite{Schmidt}, the spin-boson network model
\cite{Mandt}, and light-matter coupled systems \cite{Yam}. The simplest case
is the Jaynes Cummings model, for a single-mode cavity, containing one
two-level atom:
\begin{eqnarray}
H_{JC}&=&
\label{JC}
\omega_{0}a^{\dagger}a+\omega_{a}\sigma^{+}\sigma^{-}+\eta(a^{\dagger}\sigma^{-}+a\sigma^{+}),
\end{eqnarray}
where $\omega_{0}$ and $\omega_{a}$ are the energies of the mode and atom,
$\eta$ is the vacuum Rabi frequency that characterizes the photon-atom
interaction strength and $\sigma^{\pm}$ are the atomic raising and lowering
operators. We have written this Hamiltonian using the rotating wave
approximation (RWA), since it then commutes with the total number of
excitations, $[H_{JC},N]=0$, where $N =
n_{a}+n_{\sigma}$, $n_{\sigma}=\sigma^{+}\sigma^{-}$ is the number operator for
fermions, and $n_{a}=a^{\dagger}a$ is the number operator for bosons.  In Fig.2 we
show numerical calculations of the dynamics of the Jaynes-Cummings model
described by $H_{JC}$, preparing the system in an initial state
$\ket{\psi(0)}=\ket{\alpha_{0},e}$, with $|\alpha_{0}|^{2}$ particles in the
field mode, and the two level system in the excited state $\ket{e}$. As
expected, we find that the atomic and the field correlations change in time,
while $g_{tot}^{(2)}(t,t)$ remains constant, since the $U(1)$ symmetry is locally broken,
$[H_{JC},n_{\sigma}]\ne 0$ and $[H_{JC},n_{a}]\ne 0$. In order to lose the stationarity, it is sufficient to break the symmetry with respect to the total number
operator to lose the stationarity, for example having a coherent driving term
like $\Omega_{\sigma}(t)=h(t)(\sigma^{+}+\sigma^{-})$ or
$\Omega_{a}(t)=h(t)(a^{\dagger}+a)$ \cite{Stiev}. The Jaynes
Cummings model beyond the rotating wave approximation contains terms that break the $U(1)$ symmetry, allowing creation and
destruction of excitations. Then, when the coupling strength $\Omega$ is increased, a phase transition occurs occurs where particles are spontaneously created. This generates a dynamics of
the total SOC functions. {Similar behavior should be seen in any system which breaks
a $U(1)$ symmetry while undergoing a phase transition, for example, in the
case of Hubbard Hamiltonians inside a cavity \cite{Chen}.}

\begin{figure}
\begin{center}
  \includegraphics[width=1.\columnwidth]{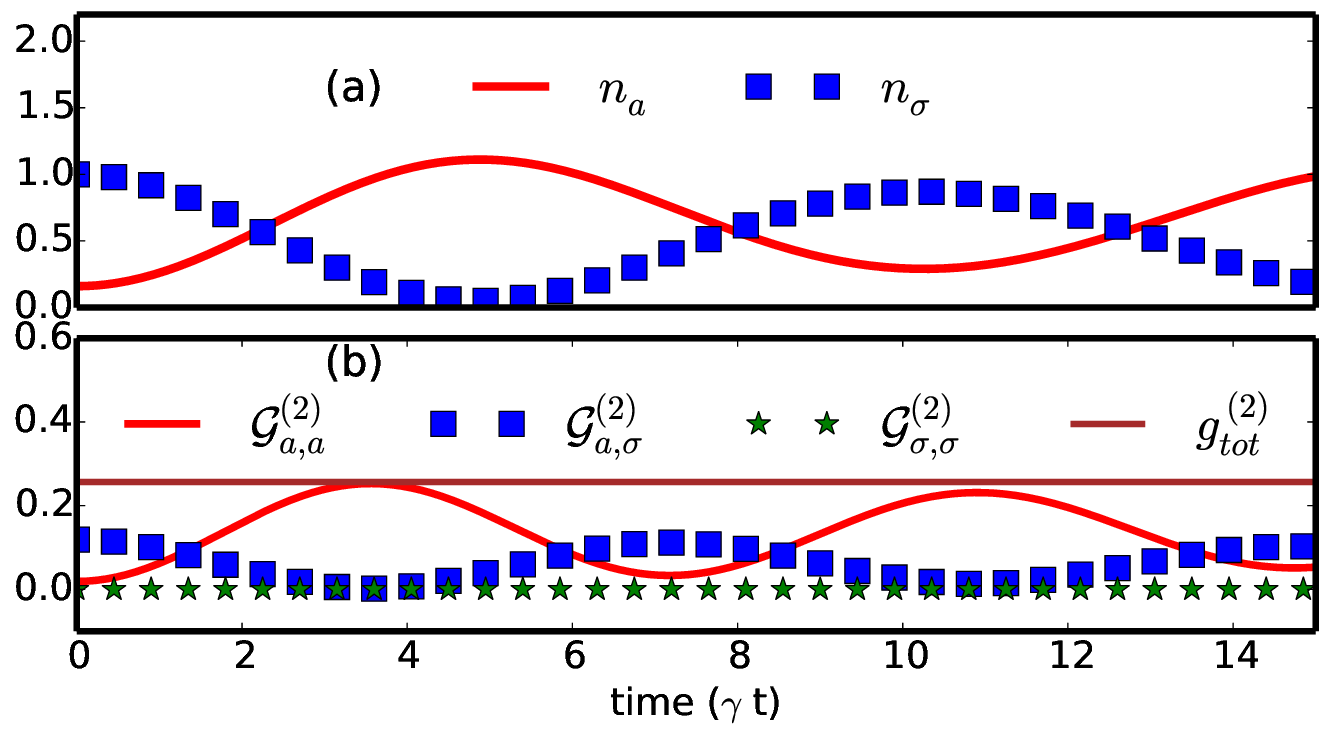}


 \label{fig:Ng2}
\caption{Numerical solution for $H_{JC}$, with a Rabi term
$\eta=0.25$ and $\omega_{a}=\omega_{\sigma}=0$, prepared in an initial state
with $|\alpha_{0}|^{2}=0.18$ particles in the cavity mode and the atom in the
excited state. (a) The number of photons, $n_a$, and excitation level of the atom, $n_{\sigma}$. $n_{\sigma}=0$ corresponds to the atom in its ground state, $n_{\sigma}=1$ to the excited state.
(b) The correlation functions $\mathcal{G}^{(2)}_{i,j}=\braket{a_{i}^{\dagger}a_{j}^{\dagger}a_{i}a_{j}}/N^{2}$ for the atomic, bosonic and for the cross terms. The total correlations
for the emitted light is a constant of motion; here it is always
sub-Poissonian ($g^{(2)}_{tot}<1$).}
\end{center}
\end{figure}
\textit{Conservation of $g_{tot}^{(m)}(t,t)$ in dissipative systems}.
We now consider dissipative systems, whose dynamics is described by a master equation in the Lindblad form\cite{Englert}:
\begin{equation}
\label{master}
\frac{\partial \rho}{\partial t}
=-i[H,\rho]+\mathcal{D}(\rho)
\end{equation}
where
\begin{equation}
\label{dissipation}
D(\rho)=\sum_{i}\frac{\gamma_i}{2} (2c_{i}^{\dagger}\rho c_{i}-c_{i}^{\dagger}c_{i}\rho-
\rho c_{i}^{\dagger}c_{i})
\end{equation} 
In this case, the time derivatives of $\braket{N}$ and $\braket{J}$ are not zero, but, from \eqref{conditcl}, the correlation function is still
conserved if
\begin{align}
\label{condit}
   \frac{d}{dt}\braket{N}&=\kappa \braket{N}\\
   \frac{d}{dt}\braket{J}&=m \kappa \braket{J}
\end{align}
where $\kappa$ is an arbitrary constant.

The time derivative of the mean value of an operator, using the Schr\"odinger 
picture, can be expressed as: 
\begin{equation}
\frac{\partial}{\partial t}\braket{A}
=\textmd{Tr}\left\{\frac{\partial}{\partial t} (A\rho)\right\}
=\textmd{Tr}\left\{A \frac{\partial \rho}{\partial t}\right\}.
\end{equation} 
Using this, with Eq.\eqref{master} for $\partial \rho/\partial t$,
the conditions for the conservation of the total correlation
function in Eq.\eqref{condit} become
\begin{align}
\label{eqop1}
-i\braket{[H,N]}+ 
\sum_{i} \frac{\gamma_i}{2}\braket{(2c_{i}^{\dagger}Nc_{i}-c_{i}^{\dagger}
c_{i}N-Nc_{i}^{\dagger}c_{i})}
&=\kappa \braket{N}
\\
\label{eqop2}
-i\braket{[H,J]}+ 
\sum_{i} \frac{\gamma_i}{2} \braket{(2c_{i}^{\dagger}Jc_{i}-c_{i}^{\dagger}c_{i}J-Jc_{i}^{\dagger}c_{i})}
&=\kappa m\braket{J}
\end{align}
In general these conditions cannot be satisfied, but if we consider the case where
all the loss rates are equal, $\gamma_i=\gamma$, the second terms on the left hand side of each equation can be evaluated to $-\gamma \braket{N}$ and $-\gamma  m \braket{J}$. Thus, with $[H,N]=0=[H,J]$ as before, and $\kappa=\gamma$, the total correlation function is conserved.

This result can be summarized in a Corollary:

\textbf{Corollary:}\textit{ For any dissipative system with an arbitrary number
of modes, described by an Hamiltonian $H$, and with a linear dissipator
$\mathcal{D}(\rho)$, where each mode decays {with the same rate $\gamma$}, iff $H$ globally possess a $U(1)$ symmetry, i.e. iff
$[H,N]=0$, where $N$ is the
total number operator, then the total $m^{th}$ order equal time 
correlation function
$g^{(m)}_{tot}(t,t)$ is a conserved quantity.}

The corollary also applies to the case of linear gain instead of loss (swapping the $c$ and $c^{\dagger}$ operators in Eq.\eqref{dissipation}), but not nonlinear dissipative processes ($c$ replaced by $c^2$ etc).

 Although the requirement for all the modes to have equal loss rates is a significant restriction, there are many physical systems made of identical 
elements, for which the losses are expected to be be equal, so the theorem applies. For examples, we can look to photonic lattice structures \cite{Oz2,Castel}, arrays of semiconductor micro cavities with same detuning \cite{Rodr} and continuous systems, like waveguide or waveguide networks with isotropic geometry \cite{Mukh}.



\textit{Single mode quantum field}. We consider first the dynamics of the SOC
function for a single mode photonic mode ($a$, $a^{\dagger}$) without any pumping term. This may be
linear, but, as previously discussed,there may be self-interaction terms, such as
a Kerr non-linearity, provided they depend only on the number operator
$a^{\dagger} a$. The conditions 
for our theorem are satisfied and the SOC is conserved during the dynamics of the system as the population decays away. However, this is not trivial, as in the case without dissipation; both $d\braket{N}/dt$
and $d\braket{J}/dt$ are non zero, but their contributions cancel. If a pump is introduced then $[H,N]\neq 0$, so the SOC will become time dependent. Hence, in an experiment with a pulsed pump, a non-classical correlation can develop while the pump is on, and this will remain unchanged as the field decays away after the pump switches off.

\textit{Two modes quantum field}. We next consider again the case of two undriven  bosonic modes, coherently
coupled, as in \eqref{Htm}, with the same linear dissipation for each mode. Using the master
equation approach it is easy to show that the dissipation terms appearing in Eqs.\eqref{twomodeG} and \eqref{twomodeN} are exactly cancelled by those in the autocorrelation functions, \eqref{twomodecross}, so, as expected, the total SOC is conserved.
However, it is instructive to look in more detail at the development of non-classical correlations in a two mode system with Kerr non-linearities.

\begin{figure}
\begin{center}
   \includegraphics[width=1.\columnwidth]{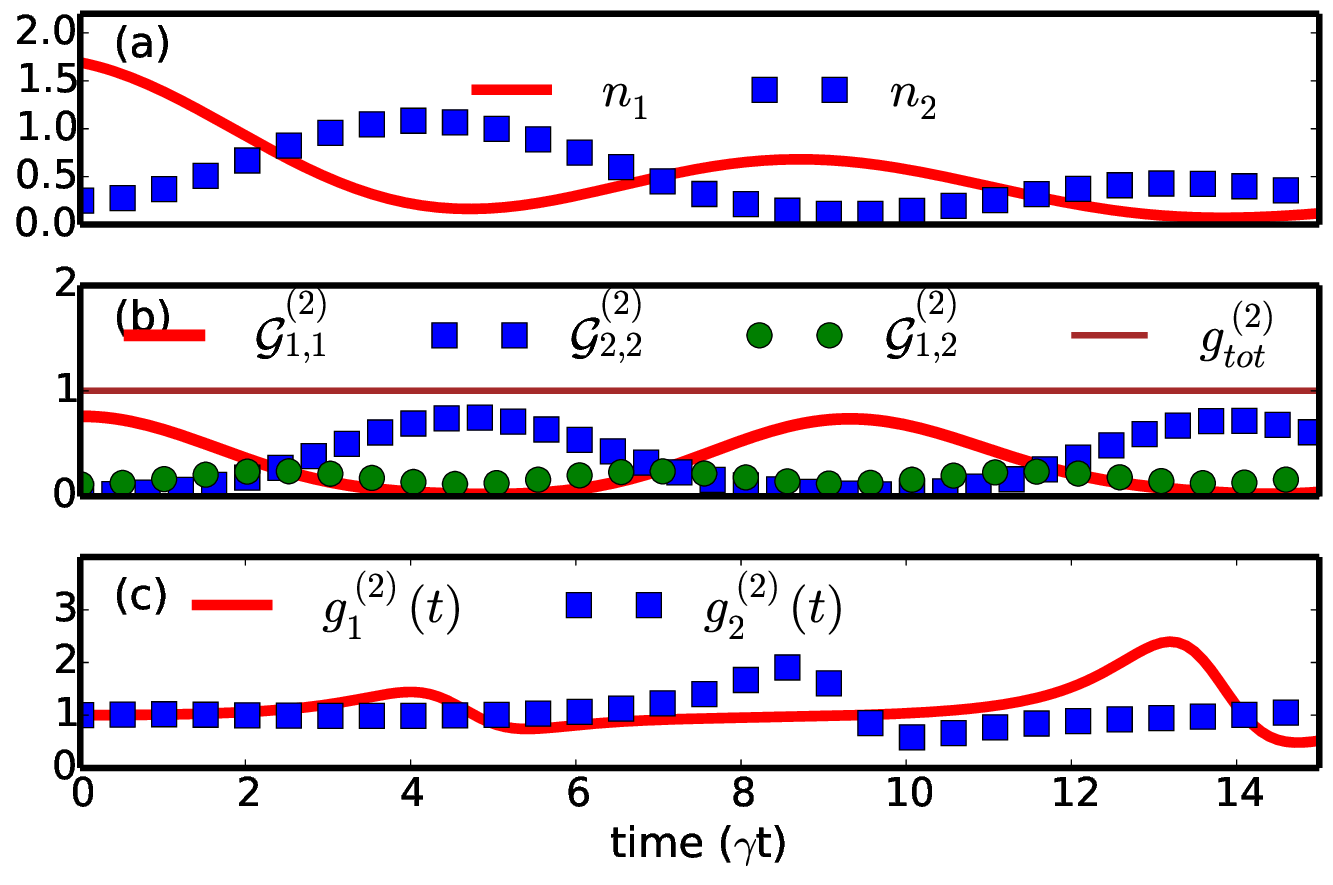}
   

  \label{fig:Ng2}
\caption{Numerical solutions for the number of excitation for two coupled bosonic modes,
with coupling $\tau=1.5 \gamma$ and nonlinearity $g=0.25 \gamma$, prepared into an
initial state with $|\alpha_{1}|^{2}=1.7$ particles in mode 1 and
$|\alpha_{2}|^{2}=0.22$ particles in mode 2 ($\alpha_1$ and $\alpha_2$ both
real and positive), and $\omega_{1}=\omega_{2}=0$. 
(a) Populations of the two modes, $n_1$ and $n_2$.
(b) The correlation functions $\mathcal{G}^{(2)}_{i,j}=\braket{a_{i}^{\dagger}a_{j}^{\dagger}a_{i}a_{j}}/N^{2}$ for each mode and for the cross terms. The total correlation function for the
emitted light is a constant of motion; in this case $g_{tot}^{(2)}=1$, as the initial state is coherent.(c) The time evolution of the normalized SOC functions for the individual modes; each shows a non-trivial behavior, even though the total correlation function is stationary.}
\end{center}
\end{figure}

We use the positive-P
representation developed by Drummond et al.\cite{Drum}, to
transform the master equation, Eq(\ref{master}), into a set of stochastic
differential equations which are solved using Monte-Carlo methods. We consider
a system of two non-linear bosonic modes described by the Hamiltonian
$H_{tm}$, with the addition of on-site non-linearities $g(a^{\dagger
2}_{1}a^{2}_{1}+a^{\dagger 2}_{2}a^{2}_{2})$,
prepared in an initial coherent state,
$\ket{\psi(0)}=\ket{\alpha_{1},\alpha_{2}}$, with $|\alpha_{1}|^{2}$ particles
in one mode and $|\alpha_{2}|^{2}$ particles in the other mode. We evaluate
the normalized second order correlation function for each mode and for the
cross-correlations. While the particles move between cavities and disappear due to dissipation, the quantities $\mathcal{G}_{i,j}^{(2)}=\braket{a_{i}^{\dagger}a_{j}^{\dagger}a_{i}a_{j}}/N^{2}$, describing the correlation functions normalized to the total number of particles, undergo a dynamical evolution.
However, it can be clearly seen that the total correlation function
is constant in the whole time interval, as our theorem requires. 
The conserved $g^{(2)}_{tot}=1$, as the initial state is coherent. By contrast,if we look at the 
the individual cavity SOCs, $g^{(2)}_1(t,t)$ and $g^{(2)}_2(t,t)$, non-classical behaviour is apparent, with their values falling below one at some stages of the evolution. This demonstrates a situation where the total emission has classical statistics, but quantum effects can be observed if the individual modes are resolved.

\textit{Continuum limit}. Until now we have considered a discrete
collection of quantum systems. However, our  \textbf{Theorem} can be extended to the case of a system with a continuum of modes, for example, the
optical field in a waveguide. If the Hamiltonian commutes with the
total number operator\cite{Distrib}, 
\begin{align}
{N}=\int dx \, a^{\dagger}(x)a(x),
\end{align}
then the total SOC, 
\begin{align}
g_{tot}^{(2)}
=\frac{1}{{ N}^2} 
\int^{+\infty}_{-\infty} dx \, dx'\,
a^{\dagger}(x) a^{\dagger}(x') a(x) a(x')
\end{align}
is stationary, provided any losses are linear and independent of $x$. This is true even when individual
point-like components, $g^{(2)}(x,x')$, experience temporal dynamics.

\textit{Conclusion}. In conclusion we have shown that the $m^{th}$ order
quantum correlation function is a constant of motion, for systems where
the  Hamiltonian possess global $U(1)$ symmetry and any dissipation is identical for each mode of the system and linear in the system operators. For a multimode system,
the $m^{th}$ order quantum correlation function may change dynamically in each mode, due to the their mutual interactions. However, total correlation function will 
still be a constant of motion. 
Our theorem suggests that the total correlation function may be an interesting
parameter to measure in systems which undergo phase transitions characterized
by the breaking of a global $U(1)$ symmetry.
Our results also demonstrate that care is necessary when using the SOC as a probe for
non-classical physics: in systems with $U(1)$  symmetry, it may be necessary to
isolate light from individual modes, rather than looking at the statistics of
the total emitted light.

We gratefully acknowledge support by EPSRC via
Programme Grant No. EP/N031776/1.
\bibliographystyle{ieeetr}
\bibliography{pap}

\end{document}